\documentclass{jetpl}
\twocolumn

\usepackage{caption}
\usepackage{graphicx}
\usepackage{subcaption}
\usepackage{cite}
\usepackage[dvips]{graphicx}
\usepackage{multirow}
\DeclareGraphicsExtensions{.pdf,.png,.jpg,.eps,.ps}
\RequirePackage{caption}
\DeclareCaptionLabelSeparator{point}{. }
\makeatletter
\newcommand{\keepwithnext}{\@beginparpenalty 10000}
\makeatother

\title{The role of photon fields of local sources in the formation of the cosmic rays energy spectrum}
\rtitle{The role of photon fields of local sources in the formation of the cosmic rays energy spectrum}
\sodtitle{The role of photon fields of local sources in the formation of the cosmic rays energy spectrum}

\author{S.~E.~Pyatovsky~\thanks{~e-mail: vgsep@ya.ru}~\textsuperscript{a,~b},~S.~K.~Machavariani~\textsuperscript{a}}

\rauthor{S.~E.~Pyatovsky, S.~K.~Machavariani}
\sodauthor{S.~E.~Pyatovsky, S.~K.~Machavariani}

\address{\textsuperscript{a}~Lebedev Physical Institute, Russian Academy of Sciences, Moscow, Russia}
\address{\textsuperscript{b}~Moscow Aviation Institute (National Research University), Moscow, Russia}

\abstract{The influence of photon fields of local sources (LS) on the formation of the cosmic ray (CR) energy spectrum ($E_0$) is considered. A model for the formation of irregularities in the CR spectrum, based on the local source model, is proposed. It is shown that the CR spectrum can be formed as a sum of partial spectra formed near local CR sources during the passage of primary cosmic radiation (PCR) through the photon fields of the sources. Local sources can be the most active stars of various types of variability in the stellar population, - from dwarfs to hypergiants. Studies have shown that the so-called "bump" near $E_0=100~PeV$ is formed in the photon fields of Mira-type variable stars, which are typically multiple stellar systems, consisting of a red giant and a white dwarf.\newline
\newline
}


\begin{document}

\maketitle
{\bf 1~Introduction.}\\ \keepwithnext
The causes and types of irregularities in the $E_0$ spectrum of PCR remain the subject of scientific discussions and modeling. In particular, questions remain regarding the energies at which knees should appear in the spectra of light (protons and $He$) and heavy ($>He$) nuclei in the mass composition of the PCR. Answers to these questions help us understand the sources and mechanisms of PCR acceleration.

Modeling of the CR spectrum as a function of $E_0$ in accordance with the local-source model was proposed by A.~M.~Hillas in 1979 and studied by S.~I.~Nikolsky~\cite{1} in 1988. Calculations using the LS model were performed by S.~E.~Pyatovsky and A.~D.~Erlykin~\cite{2} in 1995. Later, the LS theory was developed in the works of A.~D.~Erlykin and A.~V.~Wolfendale~\cite{3,4}. In particular, a comparison between the results of studies of the CR mass composition in the knee region of the CR spectrum at $E_0=3-5~PeV$, carried out by A.~D.~Erlykin, A.~V.~Wolfendale and V.~P.~Pavlyuchenko~\cite{5,5-1,5-2}, and the results of the PAMIR experiment using X-ray emulsion chambers (REC) showed~\cite{6} that the local source \textrm{$\gamma$}~Vel (a multiple Wolf-Rayet star system in the constellation Vela) cannot be regarded as the sole source responsible for forming the CR spectrum in the knee and bump regions.

Calculations performed in accordance with the LS model showed~\cite{2} that the main part of CR is secondary in relation to PCR, and that the mass composition of CR is formed mainly during the passage of primary CR through the LS photon field in the process of photo-disintegration of nuclei. According to the local source model, a stream of nuclei moves away from the LS following the expanding photon field of the star, while its mass composition remains unchanged up to energies of $E_0\sim0.1~PeV$. At these energies, photo-nuclear and photomeson reactions begin, and the first irregularities in the PCR spectrum are observed, for example, the so-called "early knee" in the CR energy spectrum at $E_0=0.1~PeV$.

\begin{figure*}[h]
\centering
\includegraphics[width=\linewidth]{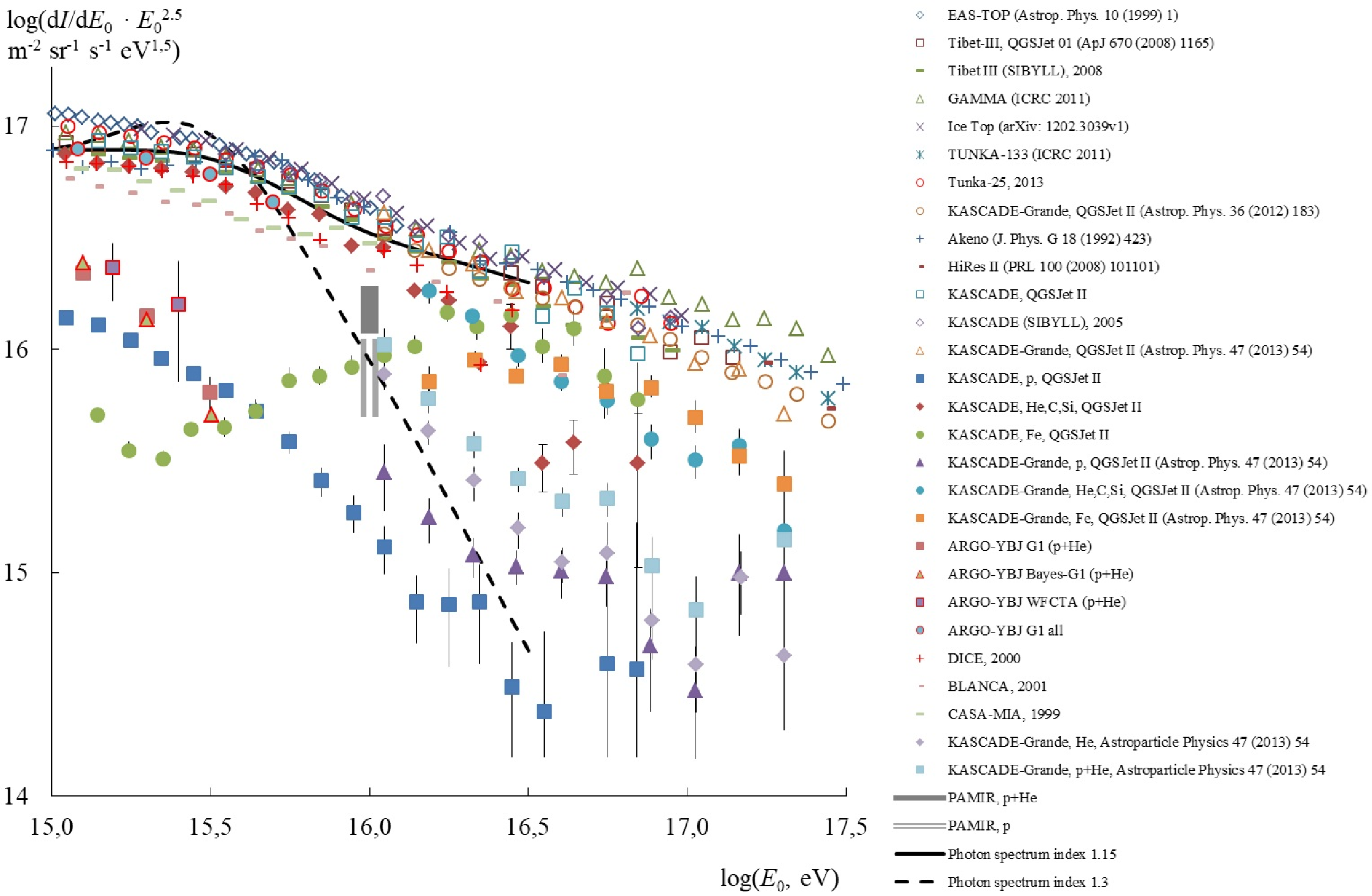}
\caption{Differential spectra of the CR nuclei as a function of $E_0$ in the knee region at $E_0=3-5~PeV$~\cite{6,7,8,9}.}
\label{fig1}
\end{figure*}

Figure~\ref{fig1} shows the spectra of light and heavy nuclei in the CR mass composition as a function of $E_0$ in the energy range of $1-100~PeV$. Together with data from experiments studying the characteristics of extensive air showers (EAS) (EAS-method), Figure~\ref{fig1} also presents data from the PAMIR experiment~\cite{6,6-1}, which used the REC-method, making it possible to study the characteristics of EAS cores within a radius of up to several tens of centimetres from the EAS axis with a spatial resolution of $\sim30~\mu m$. In particular, based on data from the PAMIR experiment, the "halo-method" \cite{6} was developed using exclusively experimentally obtained statistics of $\gamma$-quanta families with a halo (the so-called "halo"). This is the only method for estimating the mass composition of CRs at $E_0>0.1~PeV$ that is weakly dependent on models of EASs propagation through the atmosphere. The method makes it possible to analyse events in EAS cores formed by protons and $He$ nuclei during the passage of the EASs generated by them through the atmosphere, while information about the interaction of the primary nucleus with air atoms remains minimally distorted.

At present, it is considered most probable that the acceleration of PCR to high and ultra-high energies $E_0>1-100~PeV$ occurs at the shock wave fronts of supernovae or in the magneto-spheres of pulsars, since these mechanisms can theoretically provide the energy required to sustain the experimentally observed CR energy-flux density in the Galaxy. The resulting CR spectrum has a differential index of $\gamma\sim2.5-2.7$ in the relatively flat part of the spectrum. However, calculations show that hydro-magnetic acceleration at shock fronts provides a maximum energy of PCR nuclei of no more than $0.2-0.3~PeV$, while a further increase in magnetic field strength near the stellar surface to the maximum observed values of $0.1-0.2~GOe$ leads to an increase in synchrotron losses of nuclei, thereby limiting the maximum energy of CR nuclei accelerated at supernovae shock fronts to $\sim100~PeV$.

At the same time, magnetic field values reaching tens of gauss are typical for red giants. For example, the magnetic field induction of the star \textrm{$\epsilon$}~Leo in the constellation Leo on the red giant branch is $B=+49\pm6~G$~\cite{10}, indicating dynamo-processes in the although convective shell of the star. Despite the fact that stars with magnetic fields of up to kG are also observed~\cite{11}, these stars are mainly dwarfs, and their total energy output is insufficient to explain the CR spectrum at energies above $>100~PeV$.

\begin{figure}[h]
\centering
\includegraphics[width=\linewidth]{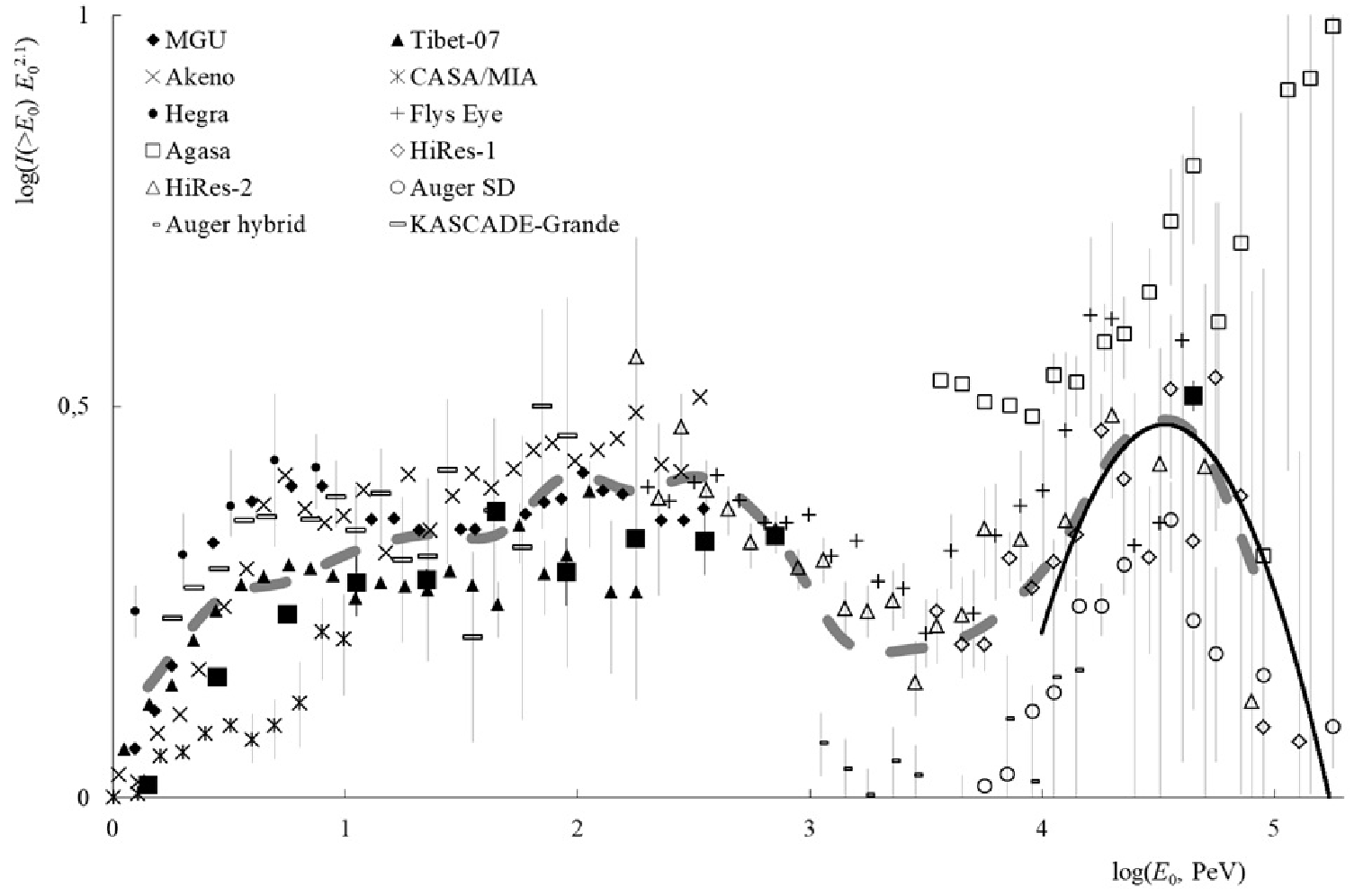}
\caption{Integral CR spectrum at $E_0>1~PeV$ with marked averaging (dashed line)~\cite{13}.}
\label{fig2}
\end{figure}

Thus, acceleration of PCR at supernovae shock waves is possible, but not to the ultra-high energies of CR nuclei. In addition, at $E_0>0.1~PeV$, so-called "irregularities" begin to appear in the CR spectrum, such as an early knee around $0.1~PeV$, a knee and a bump at $3-5~PeV$ and around $100~PeV$, the irregularities at $10^{18}-10^{20}~eV$, and others, as shown in Figure~\ref{fig2}. All of the above indicates that there is no single mechanism responsible for the formation of the CR spectrum. The question of the existence of the GZK-cutoff in the CR spectrum at $E_0\sim10^{20}~eV$~\cite{11-1}, caused by the interaction of protons with photons of the relic radiation, is also being discussed. For example, one of the tasks of the LHAASO experiment~\cite{12} is to detect neutrinos correlated with the diffuse radiation of the Galaxy, in order to obtain data related to the physics of $\Delta$-resonances in hadronic processes.

Another more "active" group of stars consists of variable stars. The role of variable stars in the formation of the CR spectrum was studied in~\cite{13}. Unlike stars with constant apparent brightness, variable stars change their brightness as a result of non-stationary processes. For example, in novae and supernovae within multiple systems, variability may be associated with the accretion of matter and with the emergence of strong magnetic-field structures at the stellar surface, for example, when the axes of stellar rotation and magnetic fields do not coincide.

\begin{figure}[h]
\centering
\includegraphics[width=\linewidth]{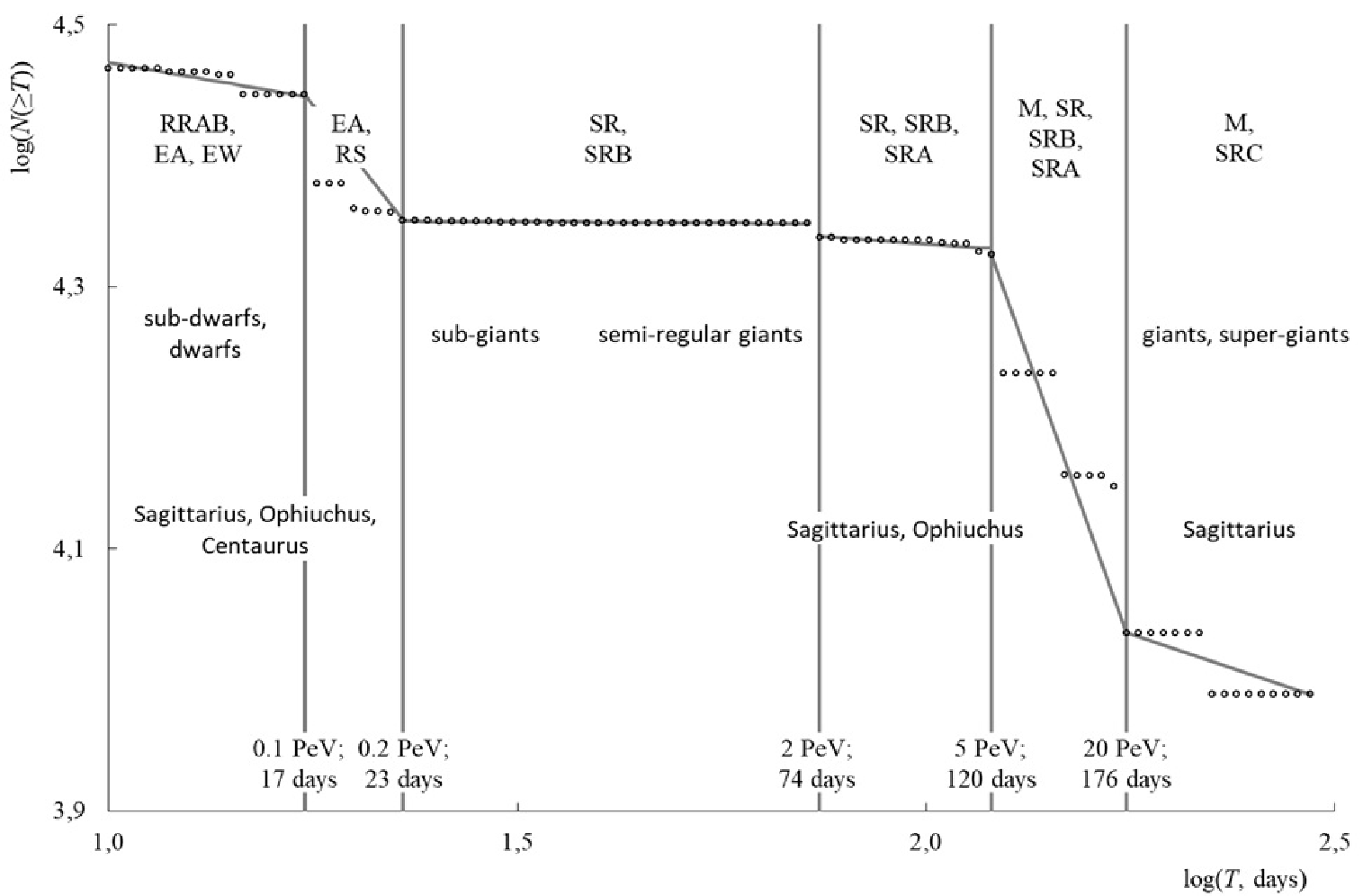}
\caption{Integral distribution of variable stars as a function of their logarithmic periods~\cite{13}. The dominant types of stars and constellations for different maximum CR energies $E_{0max}$ are shown.}
\label{fig3}
\end{figure}

Figure~\ref{fig3} shows the integral spectrum of variable stars depending on their logarithmic periods. As shown in~\cite{13}, sources of low-energy PCR at $E_0<0.1~PeV$ can be variable dwarfs located mainly in the constellations Sagittarius, Ophiuchus, and Centaurus; at intermediate energies, $E_0=0.2-2~PeV$, they can be variable subgiants and giants from Sagittarius and Ophiuchus; and at high energies, $E_0>5~PeV$, they can be variable giants, supergiants and hypergiants from Sagittarius.

{\bf 2~Local source model conditions.}\\ \keepwithnext
The goal of the local source model is to describe the CR spectrum over the widest possible energy range and to answer the question of which stars can be sources of PCR of high and ultra-high energies.

In the LS model, it is assumed that a power-law spectrum by $E_0$ over the entire range of nuclear energies should be formed near the PCR source by the magnetic fields present in LS, without any knees or bumps in the PCR spectrum. Local sources are active stars with various types of variability: eruptive (for example, the flaring multiple star system Luyten 726-8 in the constellation Cetus), pulsating (for example, the Mira $\textrm{o}$~Cet system consisting of a red giant and a white dwarf in Cetus) and explosive variable stars. The greater the age and metallicity of a star, the heavier the mass composition of the PCR it forms, ranging from hydrogen- and helium-burning products to $Fe$. In addition, stars in the final stages of their evolution have stronger magnetic fields, which are necessary for accelerating the PCR to high and ultrahigh energies.

In the local source model, it is assumed that PCR protons lose energy in photomeson reactions such as $\gamma(p,n)\pi^+$ and $\gamma(p,p)\pi^0$, with cross section $\sigma_{\gamma p}$ and known decay-channel probabilities. The nuclei of PCR formed in LSs (stars of various types of variability), while passing through photon fields near the LSs, form local spectra with a maximum energy $E_0$ in the spectrum $E_{0max}$~\cite{2}. In~\cite{13} the relationship between the period $T$ of a variable star serving as a PCR source and $E_{0max}$ was established, and it was also shown that $T\sim E_{0max}$. It is assumed that the superposition of local PCR spectra formed near PCR sources produced the total CR spectrum.

The local PCR source is surrounded by the photon field of the star, with a photon energy spectrum described by a power law:

\begin{equation}
\frac{dn}{d\epsilon}\sim\epsilon^{-\alpha}~photon\cdot cm^{-3}\cdot MeV^{-1}
\label{eq1}
\end{equation}

with the maximum photon energy $\epsilon_{max}$ in the photon field spectrum and the optical thickness $h$ of the photon field.

The interaction of a PCR nucleus with the photon field~(\ref{eq1}) of the LS leads to photodisintegration into a proton and a nucleus with mass number $A-1$ with the probability of such a disintegration being one. Taking this process into account, the system of differential kinetic equations describing changes in the mass composition of PCR nuclei as they pass through the photon field of the LS is written as follows: nuclei with mass number $A$ are removed according to the cross section $\sigma_A(E_n)$, while nuclei with mass number $A+1$ become sources of CR:

\begin{equation}
\frac{\partial}{\partial h}\frac{dI_A}{dE_n}=-\frac{1}{\lambda_A(E_n)}\frac{dI_A}{dE_n}+\frac{1}{\lambda_{A+1}(E_n)}\frac{dI_{A+1}}{dE_n}
\label{eq2}
\end{equation}

where $dI_A/dE_n$ are the differential intensities of nuclei with energy $E_n$ per nucleon of the nucleus, $\lambda_A(E_n)$ are ranges of nuclei with mass numbers $A$ in the LS photon field~(\ref{eq1}).

For the differential intensity of nuclei with energy per nucleus $E_A\equiv E_0$, equations~(\ref{eq2}) are rewritten after transformations $E_n=E_A/A$ и $E_A=E_{A+1}-E_{A+1}/(A+1)=A/(A+1)E_{A+1}$.

\begin{figure}[h]
\centering
\includegraphics[width=\linewidth]{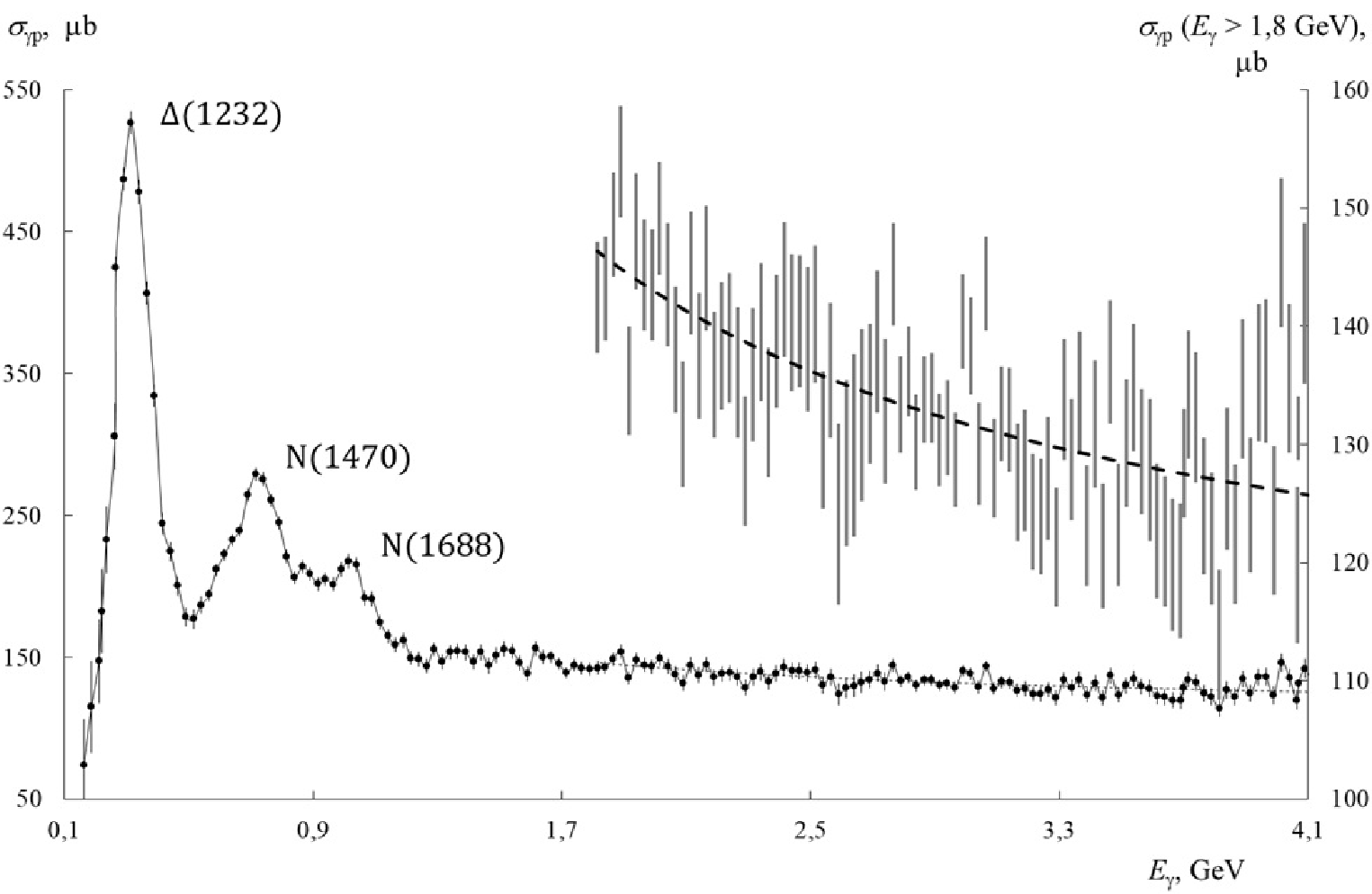}
(a)
\includegraphics[width=\linewidth]{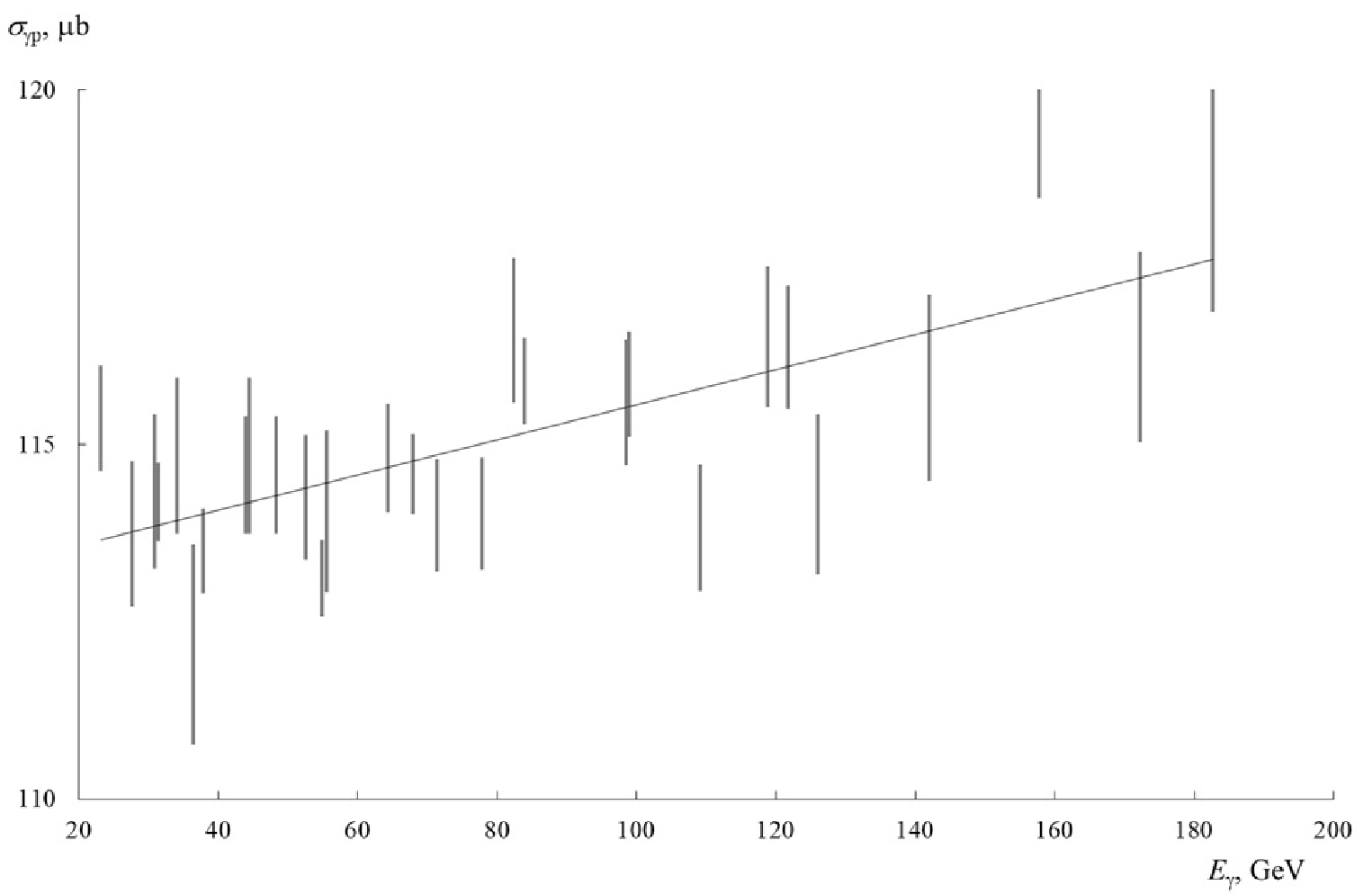}
(b)
\caption{Dependence of the total hadron cross section of the $\sigma_{\gamma p}$ $\gamma$-proton interaction on the $\gamma$-quantum energy $E_\gamma$~\cite{14,15,16,17}. The dotted line is the approximation~(\ref{eq5}) of $\sigma_{\gamma p}$ for the interval of $E_\gamma=1.8-4.0~GeV$ (right axis)~(a); the growth of $\sigma_{\gamma p}$ in the range of $E_\gamma=20-200~GeV$ as $\sigma_{\gamma p}\sim(23\pm5)10^{-3}E_\gamma$~(b).}
\label{fig4}
\end{figure}

The differential intensity of protons in CR, formed after the passage of PCR through the photon field of the LS, is determined by two sources of secondary protons. The first source consists of secondary protons produced because of the photodisintegration of nuclei~(\ref{eq2}). The second source consists of higher-energy protons that have lost part of their energy in photomeson reactions $\gamma(p,n)\pi^+$ and $\gamma(p,p)\pi^0$ through the decay of $\Delta$- and nucleon resonances, as illustrated in Figure~\ref{fig4}, which shows the cross sections of photomeson reactions of protons depending on the energies of $\gamma$-quanta:

\begin{equation}
\begin{split}
\frac{\partial}{\partial h}\frac{dI_p}{dE_n}=-\frac{1}{\lambda_p(E_n)}\frac{dI_p}{dE_n}+\frac{1}{\lambda_p(\frac{E_n}{1-K})}\frac{dI_p}{d(\frac{E_n}{1-K})}+\\
+\sum\limits_{A=2}^{56}\frac{1}{\lambda_A(E_n)}\frac{dI_A}{dE_n}
\end{split}
\label{eq3}
\end{equation}

where for protons $E_n\equiv E_0$, $K(E_n)$ is the coefficient of inelasticity of protons in photo-meson reactions and, accordingly, $1-K$ is the fraction of energy carried away by the leading proton.

The ranges $\lambda_A$ of the nuclei in~(\ref{eq2}) and~(\ref{eq3}) for the photon fields~(\ref{eq1}) of local sources are determined by the expression:

\begin{equation}
\begin{split}
\lambda_A^{-1}(E_n)=0.5\gamma_A^{-2}\int\limits_{E'}^{2\gamma_A\epsilon_{max}}E^*\sigma_A(E^*)\,dE^*\\
\int\limits_{E^*/(2\gamma_A)}^{\epsilon_{max}}\frac{dn(\epsilon)/d\epsilon}{\epsilon^2}\,d\epsilon
\end{split}
\label{eq4}
\end{equation}

where $\gamma_A$ is the Lorentz-factor of the nucleus, $E^*$ and $E'$ are the photon energy and the threshold of the photonuclear reaction in the rest system of the nucleus, $\sigma_A(E^*)$ is the cross section of the photonuclear reaction. When calculating the interaction length of nuclei $\lambda_A$, the threshold energy is taken to be equal to the binding energy threshold of the deuteron $2~MeV$; the ranges of protons $\lambda_p$, the energy for the onset of photomeson reactions is taken to be $155~MeV$.

The range $\lambda_p$ of a proton in the photon fields of local sources is calculated similarly to expression~(\ref{eq4}), but instead of $\sigma_A(E^*)$, the cross sections of photomeson reactions $\sigma_{\gamma p}(E_\gamma)$ are taken, presented in Figure~\ref{fig4}, where the values of $\sigma_{\gamma p}$ are given for protons in the studied energy range $E_\gamma=0.1-4.0~GeV$. In particular, for the range $E_\gamma=1.8-4.0~GeV$, the approximation was used:

\begin{equation}
\sigma_{\gamma p},\mu b=(109\pm3)+(67\pm7)/E_\gamma, R^2=0.5
\label{eq5}
\end{equation}

with an asymptote of $\sigma_{\gamma p}=(109\pm3)~\mu b$, which does not contradict other studies~\cite{14}, where the asymptotic value of $\sigma_{\gamma p}=(91\pm6)~\mu b$ is presented with a correction of $10~MeV$, taking into account the change in the energy of the recorded beam scattering at $E_\gamma<2~GeV$.

The position of the first resonance peak is close to $E_\gamma=0.31-0.32~GeV$, which corresponds to the $\Delta$-resonance with the mass $\sqrt{s}=\sqrt{(\Sigma E_i)^2-(\Sigma p_ic)^2}$. Deviations of $\sigma_{\gamma p}$ from the average values in the range $E_\gamma>1.8~GeV$ remain within the statistical errors, which range from 1 to 8~$\mu b$ for the $1\sigma$ level. At higher energies $E_\gamma>20~GeV$, the $t$-value is about $5$, which allows to conclude that there is a statistically significant increase in $\sigma_{\gamma p}$ up to $\cong120~\mu b$ in the range of $E_\gamma=20-200~GeV$.

The cross sections of photonuclear reactions are calculated under the assumption of isolated resonances:

\begin{equation}
\frac{\sigma_0(A)}{\sigma_A(E^*)}=1+\frac{1}{(8~MeV)^2}\left(\frac{E^{*2}-E_0^{'2}(A)}{E^*}\right)^2
\label{eq6}
\end{equation}

where $8~MeV$ is the resonance width at half maximum, $\sigma_0\sim A$ and the resonance energy $E_0'\sim A^\beta$.
\newline

{\bf 3~The simulation results.}\\ \keepwithnext
The CR spectrum as a function of $E_0$ is calculated for different parameters of the photon fields surrounding local sources and is shown in Figure~\ref{fig1}. A high sensitivity of the CR spectrum by $E_0$ to the parameters of the photon energy spectrum~(\ref{eq1}) was obtained. The simulations showed that in the knee region at $E_0=3-5~PeV$, it is possible to obtain a local bump and a rapidly decreasing CR intensity above the knee for certain values of $E_{0max}$ (for example, the dotted line in Figure~\ref{fig1} for a photon field spectrum index $\alpha=1.3$).

The studies have shown that different characteristics of photon fields of local sources allow the formation of local CR spectra within limited $E_{0max}$ energy ranges characteristic of different types of stars. Stars with different types of variability~\cite{13}, which are local sources of PCR, have individual photon fields parameters and energy spectra $dn/d\epsilon$, characterized by different optical thicknesses, spectral indices and other parameters depending on the types of variable stars, form a range of the CR spectrum by $E_0$, characteristic of each type of star.

\begin{figure}[h]
\centering
\includegraphics[width=\linewidth]{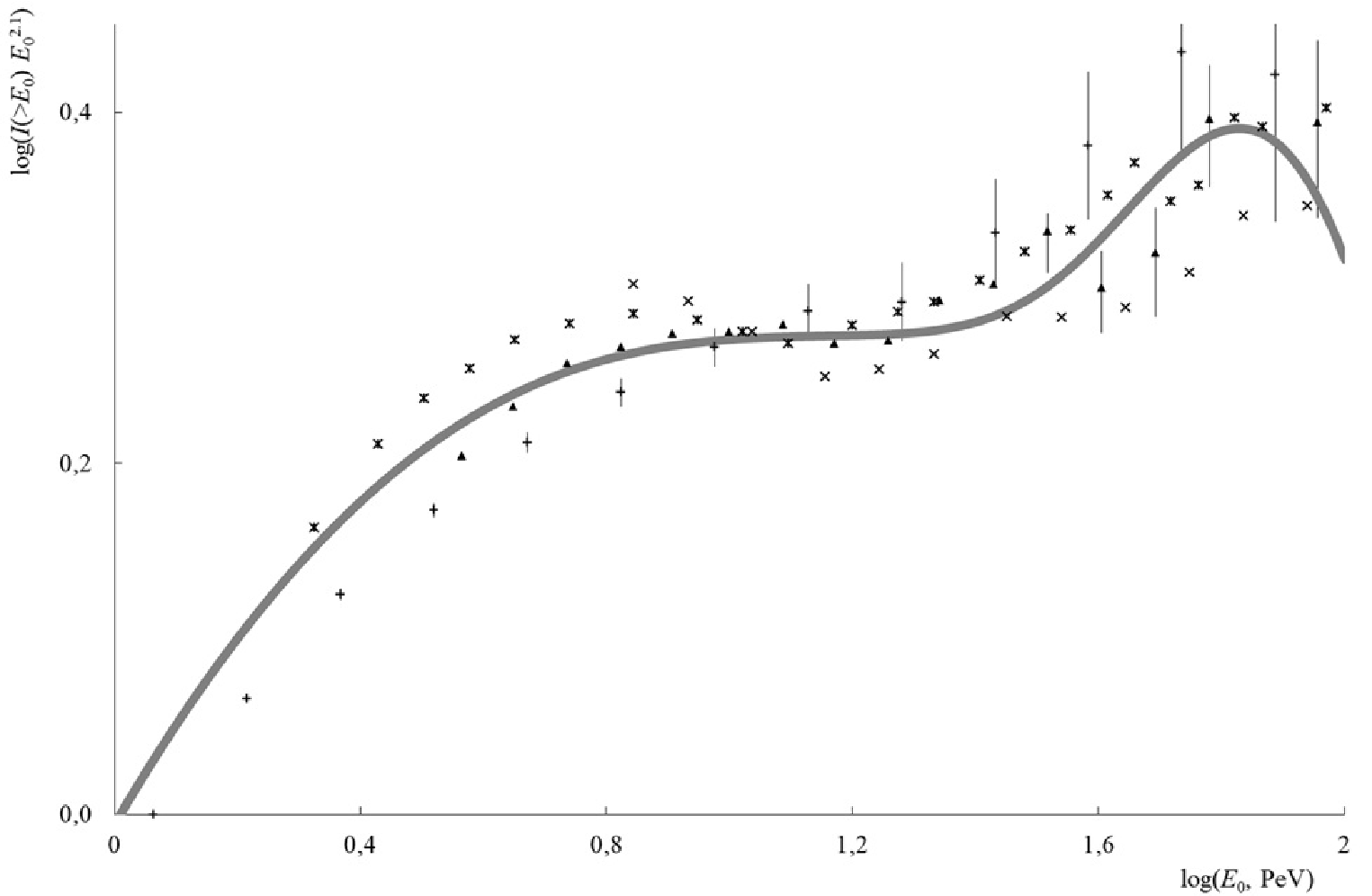}
\caption{Comparison of the spectra of the CR by $E_0$ obtained in the GAMMA, Tunka-133, Ice-Top and KASCADE-Grande experiments with the spectrum obtained using the local source model for active stars of the SRA, SRB, SRC and M type variability (gray curve)~\cite{13}.}
\label{fig5}
\end{figure}

This made it possible to describe irregularities in the CR spectrum starting from $E_0\sim0.1~PeV$ (variable red dwarf stars, whose contribution to the CR flux was studied by Yu.~I.~Stozhkov~\cite{18}), including the knee at $E_0=3-5~PeV$ and a bump with a maximum at $E_0$ around $100~PeV$, as shown in Figures~\ref{fig2} and~\ref{fig5}.

In Figure~\ref{fig5}, the CR spectrum as a function of $E_0$ is shown in the range $E_0=1-100~PeV$ as a comparison between experimental and model-obtained spectra. The maximum of the model CR spectrum bump occurs at $E_0$ around $70~PeV$. As shown in~\cite{13}, this spectrum bump is formed predominantly by the photon fields of Mira-type variable stars.
\newline

{\bf 4~Discussion.}\\ \keepwithnext
The local source model is based on several natural assumptions. The first is that the sources of CR are stars. The more active the star, - or, in many cases, the stellar system, - the higher the energies to which it can accelerate the PCR nuclei. At the same time, recent studies suggest that supernovae alone are insufficient to explain the full CR spectrum. Moreover, supernovae are not the only active objects in the Galaxy in terms of energy release and magnetic fields. For example, these could be binary and triple systems of explosive and nova-like stars with regular outbursts, in which the stellar brightness increases tenfold or more and which consist of a massive white dwarf and a red giant.

The second assumption is that the mechanisms accelerating PCR near all stars are fundamentally similar. These mechanisms provide regular CR spectra with an approximately constant spectral index $\gamma$. However, above a threshold energy $E_0$ the CR spectrum becomes irregular. This may occur when an already formed spectrum is affected by a common physical factor whose parameters differ from one type of star to another. Such a factor could be the occurrence of photonuclear and photomeson reactions taking place in the photon fields surrounding the stellar systems. Interactions between nuclei and the photon fields of local sources determine an energy $E_{0max}$, characteristic of a stellar system with a given type of variability and multiplicity. In particular, the interaction of protons with the relic radiation determines the maximum energy $E_{0max}$ in the CR spectrum, called the GZK-cutoff. A similar cutoff should also exist for nuclei when PCR nuclei pass through the LS photon field, forming $E_{0max}$.

From the above, it follows that the shape of the CR spectrum depends on the stellar population of the Galaxy. For example, if red dwarfs, which are primarily responsible for accelerating the PCR nuclei to low energies~\cite{18}, are excluded from consideration the shape of the CR spectra changes.

The same reasoning also implies an irregular mass composition of CR as a function of $E_0$. In particular, the question of whether the CR mass composition becomes heavier or lighter after the knee and the bump at $E_0=3-5~PeV$ remains under discussion, as shown in Figure~\ref{fig1}. At the same time, the LS model assumes that the CR mass composition is not constant as $E_0$ changes, which is a consistent with experimental observations.

When a stellar system with a particular type of variability experiences a multiple increase in brightness (energy release), an ejection of PCR nuclei with energies characteristic of that star may occur. For example, a star of the EA+SRC type with a registered period of $20$~years contributes to a bump in the spectra CR at $E_0=60~EeV$; a triple star system with a period of $33$~years provides a bump in the spectra CR at $E_0=180~EeV$. This also suggests that the mass composition of CR may change over time, and the higher $E_0$, the greater the variation in mass composition due to significantly different numbers of stars of a certain type, such as giants and super-giants.
\newline

{\bf 5~Conclusions.}\keepwithnext
\begin{enumerate}
\item Irregularities in the PCR spectrum begin at energies when photonuclear reactions begin. When PCR nuclei pass through the photon fields of active stars with different types of variability, CR fluxes are formed in energy ranges $E_0$ characteristic of those types of stars.

\item Each energy range of the CR spectrum as a function of $E_0$ is formed by active stars with different types of variability, from dwarfs to hypergiants. Irregularities in the PCR spectrum are produced by stars (local sources) with different characteristics of the surrounding photon fields. Consequently each energy range of the CR spectrum is characterised by its own CR mass composition, which may also very over time.

\item The local-source model does not require the development of exotic mechanisms for accelerating PCR and is based on experimentally observed parameters of stellar systems. In addition, the LS model describes the CR spectrum over the entire energy range.

\end{enumerate}


\begin{thebibliography}{99}
\bibitem{1}
S.~Karakula, S.~I.~Nikolsky, J.~N.~Stamenov, W.~Tkaczyk. The spectra of cosmic ray nuclei from compact object // $5^\textrm{th}$ Int. Symp. Very High Energy Cosmic Ray Interact. - Lodz, 1988, p.~364.

\bibitem{2}
A.~D.~Erlykin, S.~E.~Pyatovsky. Description of the knee in the energy spectrum of primary cosmic rays in accordance with the compact source model // Yadernaya Fisika, 1995, v.~58, N~4, p.~682.

\bibitem{3}
A.~D.~Erlykin, A.~W.~Wolfendale. Structure in the cosmic ray spectrum: an update // J. Phys. G: Nucl. Part. Phys. 27, 1005 (2001).\\https://dx.doi.org/10.1088/0954-3899/27/5/305

\bibitem{4}
A.~D.~Erlykin, S.~K.~Machavariani. On the Origin of the Knee in the Cosmic Ray Energy Spectrum // Bull. Lebedev Phys. Inst. 48, 31~–~34 (2021).\\https://dx.doi.org/10.3103/S1068335621020032

\bibitem{5}
A.~D.~Erlykin, A.~W.~Wolfendale. Fine structure in the cosmic ray energy spectrum as an approach to the problem of cosmic ray origin. I: status of the single source model of the knee at PeV energies // Proceedings of the 2nd International Cosmic Ray Workshop "Aragats 2011". – 2011. – 12 – 16~Sep. – Nor-Amberd, Armenia. – pp.~34~-~39.

\bibitem{5-1}
V.~P.~Pavlyuchenko, R.~M.~Martirosov, N.~M.~Nikolskaya, A.~D.~Erlykin. Properties of the difference method for searching for anisotropy of primary cosmic rays // Bull. Lebedev Phys. Inst. 44, 40~–~45 (2017).\\https://dx.doi.org/10.3103/S106833561702004X

\bibitem{5-2}
V.~P.~Pavlyuchenko, R.~M.~Martirosov, N.~M.~Nikolskaya, et~al. An anomaly in the properties of primary cosmic rays from the Vela cluster // Bull. Russ. Acad. Sci. Phys. 81, 413 – 415 (2017).\\https://dx.doi.org/10.3103/S1062873817040335

\bibitem{6}
S.~E.~Pyatovsky. The nature of the halo in the trunks of extensive air showers and the fraction of light nuclei in primary cosmic radiation at the $E_0=10~PeV$ (PAMIR experiment) // A dissertation for the degree of PhD, Dep. of Nuclear Phys. and Astrophys., Lebedev Physical Institute of the Russian Academy of Sciences, 2021.\\https://www.lebedev.ru/file/3555

\bibitem{6-1}
R.~A.~Mukhamedshin, V.~S.~Puchkov, S.~E.~Pyatovsky, S.~B.~Shaulov. Analysis of gamma-ray families with halos and estimation of mass composition of primary cosmic radiation at energies 1-100 PeV // Astroparticle Physics. – 2018. – 102. – pp.~32~-~38.\\https://dx.doi.org/10.1016/j.astropartphys.2018.05.005

\bibitem{7}
I.~De~Mitri on behalf of the ARGO-YBJ Collaboration. Measurement of the cosmic ray all-particle and light-component energy spectra with the ARGO-YBJ experiment // ISVHECRI 2014. – $18^\textrm{th}$ International Symposium on Very High Energy Cosmic Ray Interactions, EPJ Web of Conferences. – 2015. – 99. p.~08003.\\https://dx.doi.org/10.1051/epjconf/20159908003

\bibitem{8}
W.~D.~Apel, J.~C.~Arteaga-Velazquez, et al. KASCADE-Grande measurements of energy spectra for elemental groups of cosmic rays // Astropart. Phys. – 2013. – 47. – pp.~54~-~66.\\https://dx.doi.org/10.1016/j.astropartphys.2013.06.004

\bibitem{9}
H.~P.~Dembinski, R.~Engel, et al. Data-driven model of the cosmic-ray flux and mass composition from 10 GeV to 1011 GeV // Proceedings of the $35^\textrm{th}$ International Cosmic Ray Conference. – 2017. – PoS(ICRC2017). – 533.\\https://pos.sissa.it/301/533/pdf

\bibitem{10}
S.~Plachinda, V.~Butkovskaya, et al. Magnetic fields of red giants and supergiants: a review of spectropolarimetric observations // Acta Astrophys.Tau. 3(1), 6~-~11 (2022).\\https://doi.org/10.31059/aat.vol3.iss1.pp6-11

\bibitem{11}
J.~Morin, J.-F.~Donati, et al. Large-scale magnetic topologies of late M dwarfs // arXiv:1005.5552v2 [astro-ph.SR] 25~Jun~2010.\\https://arxiv.org/pdf/1005.5552v2

\bibitem{11-1}
G.~T.~Zatsepin, V.~A.~Kuzmin.  Upper limit of the spectrum of cosmic rays // JETP Letters, v. 4, N 3, pp. 78 – 80, 1966.

\bibitem{12}
Li~Wenlian, Huang~Tian-Qi, et al. Search for neutrino signals correlated with LHAASO diffuse Galactic emission // arXiv:2408.12123v2 [astro-ph.HE] 27~Jan~2025.\\https://arxiv.org/pdf/2408.12123v2

\bibitem{13}
S.~E.~Pyatovsky. Sources of primary cosmic rays at an energy of $\sim100~PeV$ // Bull. Lebedev Phys. Inst. 51, 1–7 (2024).\\https://doi.org/10.3103/S1068335623601024

\bibitem{14}
T.~A.~Armstrong, W.~R.~Hogg, G.~M.~Lewis, A.~W.~Robertson, G.~R.~Brookes, A.~S.~Clough, J.~H.~Freeland, W.~Galbraith, A.~F.~King, W.~R.~Rawlinson, N.~R.~S.~Tait, J.~C.~Thompson, and D.~W.~L.~Tolfree. Total Hadronic Cross Section of $\gamma$ Rays in Hydrogen in the Energy Range 0.265-4.215~$GeV$ // Phys. Rev. D~5, 1640 (1972).

\bibitem{15}
D.~O.~Caldwell, J.~P.~Cumalat, A.~M.~Eisner , A.~Lu, R.~J.~Morrison , F.~V.~Murphy, S.~J.~Yellin, P.~J.~Davis, R.~M.~Egloff, M.~E.~B.~Franklin, G.~J.~Luste, J.~F.~Martin, J.~D.~Prentice, and T.~Nash. Measurements of the Photon Total Cross Section on Protons from 18 to 185~$GeV$ // Phys. Rev. Lett., v.~40, N~19, 1978, p.~1222.

\bibitem{16}
D.~O.~Caldwell, V.~B.~Elings, W.~P.~Hesse, R.~J.~Morrison, F.~V.~Murphy, and D~ E.~Yount. Total Hadronic Photoabsorption Cross Sections on Hydrogen and Complex Nuclei from 4 to 18~$GeV$ // Phys. Rev. D, v.~7, N~5, 1973, p.~1362.

\bibitem{17}
E.~D.~Bloom, R.~L.~Cottrell, D.~H.~Coward, H.~DeStaebler, J.~Drees, G.~Miller, L.~W.~Mo, and R.~E.~Taylor. Determination of the total photon-proton cross section from high-energy inelastic electron scattering // SLAC-PUB-653, September 1969, (TH) and (EXP).

\bibitem{18}
V.~G.~Sinitsyna, V.~Yu.~Sinitsyna, Yu.~I.~Stozhkov. Red dwarf stars as a new source type of galactic cosmic rays // Astronomische Nachrichten. 2021. 342. 1-2. pp.~342~-~346.\\https://dx.doi.org/10.1002/asna.202113931

\end{thebibliography}
\end{document}